# On Satisfying the Android OS Community: User Feedback Still Central to Developers' Portfolios

Sherlock A. Licorish[1], Amjed Tahir[1], Michael Franklin Bosu[1] and Stephen G. MacDonell[1,2]
[1]Department of Information Science, University of Otago, New Zealand
{sherlock.licorish; amjed.tahir; michael.bosu}@otago.ac.nz
[2]SERL, School of Computer and Mathematical Sciences, Auckland University of Technology, New Zealand
stephen.macdonell@aut.ac.nz

**Abstract**

*End-users play an integral role in identifying requirements, validating software features' usefulness, locating defects, and in software product evolution in general. Their role in these activities is especially prominent in online application distribution platforms (OADPs), where software is developed for many potential users, and for which the traditional processes of requirements gathering and negotiation with a single group of end-users do not apply. With such vast access to end-users, however, comes the challenge of how to prioritize competing requirements in order to satisfy previously unknown user groups, especially with early releases of a product. One highly successful product that has managed to overcome this challenge is the Android Operating System (OS). While the requirements of early versions of the Android OS likely benefited from market research, new features in subsequent releases appear to have benefitted extensively from user reviews. Thus, lessons learned about how Android developers have managed to satisfy the user community over time could usefully inform other software products. We have used data mining and natural language processing (NLP) techniques to investigate the issues that were logged by the Android community, and how Google's remedial efforts correlated with users' requests. We found very strong alignment between end-users' top feature requests and Android developers' responses, particularly for the more recent Android releases. Our findings suggest that effort spent responding to end-users' loudest calls may be integral to software systems' survival, and a product's overall success.*

# 1. INTRODUCTION

The most successful software projects are those that, above all else, deliver features that are genuinely desired by and readily satisfy the projects' end-users [1]. In fact, there is an established sub-discipline of requirements engineering within which the importance of users' prioritized requirements is considered to be central to the development of a software product [1]. That said, end-user satisfaction with feature delivery may be tempered by other issues, such as project costs and risks. Thus, effective product *and* process management tend to result in projects that satisfy users.

In market-driven requirements engineering (MDRE), where software is developed for many potential users, the traditional requirements gathering process, including the negotiation of requirements with a single customer, does not apply. The Android OS belongs to the MDRE domain, since it is developed for millions of end-users. The early versions of the Android OS likely relied on internal Google teams and other market research for gathering requirements. However, it may be that subsequent releases and enhancements have benefitted from user reviews. A key challenge of such an approach is how these (potentially large numbers of) reviews are handled in the decision-making processes associated with the management of requirements. Since in MDRE there is no single individual or group of customers that can be directly identified for requirements elicitation, according to Regnell & Brinkkemper [2] MDRE requires that firms instead place significant emphasis on requirements prioritization. Managing the cost of this process during the development or extension of a product, addressing the most important requirements, and planning the release of that product, are also key issues for organizations to consider to stay competitive [2]. In fact, software organizations using MDRE are responsible for all requirements decisions and their associated risks. Thus, issues related to requirements overload, requirements ambiguity and prioritization challenges can influence the competitiveness of the organization [2].

Online application distribution platforms (OADPs) provide a related mechanism to MDRE. The recent proliferation of virtual communities has supported the deployment of apps through OADPs such as Apple's AppStore and Google Play. OADPs have become a successful channel for the distribution of software applications to a mass market of users. Users' feedback in this domain presents similar challenges to those of MDRE, such as how to prioritize requirements in order to satisfy previously unknown user groups, especially with first releases of a product. That said, OADPs offers an opportunity for subsequent releases to benefit from users' reviews, which serve to inform developers of users' desires for feature enhancements and corrections [3]. Users' reviews on OADPs also supports the evolution of a software product, given that reviews provide rich content comprising users' opinions and preferences for a software product [4]. Users' impressions while employing apps are also readily assessed through the ratings they provide on OADPs. Social media provides yet another avenue through which app reviews and users' sentiments may be tracked [5].

The usefulness of OADPs features in this regard has been established. For instance, a study by Pagano & Maalej [3] involving more than one million user reviews of 1,100 apps from the Apple AppStore observed that apps that were assigned lower ratings were indeed most in need of improvement. That said, it was also discovered that several highly-rated apps had issues that could be addressed to make them better [3]. The challenge is, therefore, how to determine the most useful reviews that will serve the purpose of optimally improving and evolving apps, considering the potentially huge numbers of app reviews at hand.

This paper examines and contributes knowledge about this issue, using the Android OS as a case study. We are interested in understanding what features are developed and/or perfected, by whom, and when, all potentially critical issues in terms of maintaining product quality and customer interest. These insights reveal the usefulness of users in the feedback loop and in improving product performance, and also support our understanding of the importance placed on user-centered design in OADPs. In this case we particularly examine the Android features that are most frequently mentioned in enhancement requests, and how Google's remedial efforts correlate with these requests. We provide insights into how end-users feedback improves software products and how developers respond to such feedback. We assess our findings in relation to previous technical opinions, and outline implications for this new form of software development wherein rapid releases are the central focus of developers in an open environment. We also outline avenues for future work.

The remaining sections of this paper are organised as follows: we provide our study background in Section 2, and outline our research questions towards the end of this section. We next provide our research setting in Section 3, before providing our results in Section 4. We then discuss our findings and consider their implications in Section 5, before considering threats to the work in Section 6. Finally, we provide concluding remarks in Section 7.

## 2. BACKGROUND

The relevance of end-users in perfecting software products through the feedback they provide cannot be overstated. Now seen as 'community members', they interact with and use software systems unearthing defects and identifying new features, as well as general improvement opportunities. Such activities have long been considered under the established research and practice sub-disciplines of user involvement, usability and human-computer-interaction [6]. The consensus in these disciplines is that involving end-users during the system development life cycle, particularly in early stages, provides critical insights into the likely context of system use. Awareness of this context then leads to the development and delivery of systems that are appreciated by users [7]. As noted above, users through their feedback thus inform improvements in the quality of the system that is delivered, reducing the likelihood of developers creating unwanted system features, leading to an increase in overall user acceptance of the software system.

Newer models of software development (and OADP in particular) enable developers to gather a wide range of end-users' feedback, which was not previously afforded. This thus creates a need for efficient means for extracting the most pertinent end-user requests, and those that are actionable [7]. To this end, research work has been recently focused on providing efficient means to address this issue. For instance, Guzman & Maalej [8] applied Natural language Processing (NLP) techniques to identify the features requested in reviews of seven apps. In their work users' sentiments about app features were also extracted, and a score computed for each sentiment. Because features are voluminous and there is the potential for similarities, Guzman & Maalej [8] applied topic modelling as a means of grouping related features into distinct clusters. Carreno & Winbladh [9] discovered topics from users' comments and, based on these topics, automatically generated requirements for future versions of apps. This work, like that of Guzman & Maalej [8], also employed a topic modelling approach. Specifically, the Aspect and Sentiments Unification (ASU) model was used to retrieve topics in end-users reviews from three applications on Google play. In order to determine the requirements that would be most useful to app developers, Chen et al. [10] built the AR-Miner tool that incorporates data mining techniques to select "informative" users' reviews on app stores. These reviews are then grouped using a topic modelling technique, before ranking is applied in the order of importance so that app developers can prioritize their work.

Apart from text mining of end-user feedback, another thread of work has focused on evaluating developer-led notes and reviews of apps posted on OADPs. These works have sought to understand and validate developers' claims and potential app issues based on technologies used. For instance, Finkelstein et al. [11] extracted features from apps' descriptions (release notes) as provided by app developers. They analyzed the relationship between apps names and the features they provide as evident in their descriptions. These authors also considered these factors in relation to apps' rankings and the popularity of apps (recorded as the number of downloads and the price of apps). Gorla et al. [12] extracted over 22,500 apps' descriptions from the Google Play Store. The app descriptions were first clustered into related themes based on topic similarity. The objective of the study was to assess apps' descriptions against the APIs they used in order to discover potential security violations in the apps. These violations may be related to the accessing of personal information such as handset location and phone numbers. Particular emphasis was placed on apps that accessed Android-based device users' personal information without stating that requirement in the app descriptions, but then invoke APIs that access such information. Beyond this work [12], overall, this issue is seen as a general concern in the app community [13]: the identification of APIs in the clusters that are contrary to the app descriptions are termed as anomalous, and many apps were found to be accessing the personal information of users without their knowledge [12].

We observe that whilst previous work has extracted requirements and problem features for fixing in future versions of apps, none of these works have sought to reconcile end-users' requests with developers' responses. While the Finkelstein et al. [11] study extracted apps' features from release notes provided by developers, this detail was not assessed against end-users' requests. Similarly, the study reported by Gorla et al. [12] checked app descriptions against their APIs' usage to evaluate compliance and potential exploitation of end-users, but no effort was committed towards assessing potential developers' responses after issues were diagnosed. While efforts to extract and prioritize reviews abound, and there is little doubt about the utility of the knowledge that is provided to developers through the use of these approaches, it remains unclear whether suggestions and recommendations made on OADPs are taken on by developers in improving product offerings. Such evidence would confirm the usefulness of users in the feedback loop and in improving product performance, but would also help us to understand the importance placed on user-centered design in OADPs.

We thus examined the Android OS to understand this issue, building on the work of Finkelstein et al. [11] and Gorla et al. [12] in checking how Android development teams are responding to requests of their users in new releases of the Android OS. We examine how customer requests for software changes and new features are being addressed by new releases of the Android OS. We address these issues by formally answering the following research questions:

**RQ1.** Do Android developers respond to end-users' requests for application improvements logged via the issue tracker?

**RQ2.** How are end-users' requests addressed over time?

## 3. RESEARCH SETTING

We used the Android OS community as our case study 'organization'. Issues identified by the Android community are submitted to an issue tracker hosted by Google[1]. Among the data stored in the issue tracker are the following details: Issue ID, Type, Summary description, Stars (number of people following the issue), Open date, Reporter, Reporter Role, and OS Version. We extracted a snapshot of the issue tracker, comprising 21,547 issues submitted between January 2008 and March 2014. We next gathered release notes from the official Android developers' portal[2]. Google provides, for most OS versions, two types of notes: *users'* and *developers'* release notes. We obtained and combined all available release notes for both users and developers for each version[3].

The issues and release notes were imported into a Microsoft SQL database, and thereafter, we performed data cleaning by executing previously written scripts to remove all HTML tags and foreign characters [14], and particularly those in the summary description and release notes, to avoid these confounding our analysis.

We next employed exploratory data analysis (EDA) techniques to investigate the data properties, conduct anomaly detection and select our research sample. Issues were labelled (by their submitters) as defects (15,750 issues), enhancements (5,354 issues) and others (438 issues). Given our goal of studying the how Android development teams respond to requests of their users in new releases of the Android OS we selected the 5,354 enhancement requests, as logged by 4,001 different reporters. Of the 5,354 enhancement requests, 577 were logged by members identifying themselves as developers, 328 were sourced from users, and 4,449 were labelled as anonymous. We examined the data of each request in our database to align these with the commercial releases of the Android OS. Its first release was in September 2008[4], although the first issue was logged in the issue tracker in January 2008. This suggests that the community was already actively engaged with the Android OS after the release of the first beta version in November 2007, with issues being reported just two months later. Given this level of active engagement, occurring even before the first official Android OS release, we partitioned the issues based on Android OS release date and major name change. So, for instance, all of the issues logged from January 2008 (the date the first issue was logged) to February 2009 (the date of an Android release before a major name change) were labelled 'Early versions', reflecting the period occupied by Android OS releases 1.0 and 1.1 which were both without formal names. The subsequent partition comprised the period between Android OS version 1.1 and Cupcake (Android version 1.5), and so on. It was straightforward to mine the release notes as these were labelled as per the individual releases. Thus, apart from data cleaning, we partitioned the release notes in line with the separation of enhancement requests above.

Table I provides a brief summary of the numbers of enhancement requests logged between each of the major releases, from the very first release through to KitKat – Android version 4.4. From column three of Table I (Number of days between releases) it can be noted that the time taken between the delivery of most of Android OS's major releases (those involving a name change) was between 80 and 156 days, with two official releases (Gingerbread and Jellybean) falling outside this range. The fourth column (Total requests logged) shows that the number of enhancement requests reported increased somewhat as the Android OS progressed, with this rise being particularly evident when the mean requests reported per day for each release is considered (refer to the values in the fifth column for details). Over the six years of Android OS's existence, on average, 2.7 enhancement requests were logged every day (median = 2.6, Std Dev = 2.1).

---
[1] https://code.google.com/p/android/issues/list
[2] http://developer.android.com/about/index.html
[3] Note that Early versions has only a developers' release note.
[4] http://android-developers.blogspot.co.nz/2008/09/announcing-android-10-sdk-release- 1.html

We employed NLP techniques to study these enhancement requests and the release notes in order to answer our two research questions. NLP techniques are often used to explore and understand language usage within groups and societies. We employed multiple techniques from the NLP space in our analysis, including corpus linguistic part-of-speech tagging (POS) [15] and computational linguistic n-gram analysis [16]. We also conducted multiple reliability checks. These methods are now described in turn.

### A. NLP Techniques

Prior research has established that noun terms in unstructured text reflect the main concepts in the subject of a clause. From a POS perspective, nouns are indeed reflective of specific objects or things[5]. In terms of linguistics, nouns often form the subjects and objects of clauses or verb phrases[6]. These and other understandings have been embedded as rules in NLP tools, including the Stanford parser which performs POS tagging [15]. We created a program that incorporated the Stanford API to enable us to extract noun phrases from our sample of enhancement requests and release notes, before counting the frequency of each noun as unigrams in the enhancement requests and release notes respectively (e.g., if "SMS" appeared at least once in 20 enhancement requests our program would output SMS = 20) [7]. We also checked for misspellings and plural forms of nouns through our program, ensuring that these were counted in our analysis. This enabled us to *investigate whether Android developers respond to end-users' requests for application improvements logged via the issue tracker*. The ranking of words in this manner draws on the computational linguistic technique of n-gram analysis. An n-gram is a continuous sequence of n words (or characters) in length extracted from a stream of elements [16]. We then investigated *how end-users' requests were addressed over time* by executing the above mentioned program against enhancement requests logged and release notes published across each release block in Table I. We performed a range of statistical tests and other reliability assessments to validate our findings, as described next.

### B. Reliability Assessment

Correlation analysis was used to triangulate the results observed visually (referring to Figures 1 and 2). In addition, three authors of this work triangulated the NLP findings by manually coding a random sample of 50 outputs from each block of enhancement requests and release notes from the POS and n-gram analyses (i.e., 1000 codes in total for the ten releases in Table I), to check that nouns were correctly

TABLE I. ANDROID OS ENHANCEMENT REQUESTS FOR MAJOR RELEASES

| Version (Release) | Last release date | Number of days between releases | Total requests logged | Mean requests per day |
|---|---|---|---|---|
| **Early versions** (1.0, 1.1) | 09/02/2009 | 451 | 173* | 0.4 |
| **Cupcake** (1.5) | 30/04/2009 | 80 | 64 | 0.8 |
| **Donut** (1.6) | 15/09/2009 | 138 | 141 | 1.0 |
| **Éclair** (2.0, 2.01, 2.1) | 12/01/2010 | 119 | 327 | 2.8 |
| **Froyo** (2.2) | 20/05/2010 | 128 | 349 | 2.7 |
| **Gingerbread** (2.3, 2.37) | 09/02/2011 | 265 | 875 | 3.3 |
| **Honeycomb** (3.0, 3.1, 3.2) | 15/07/2011 | 156 | 372 | 2.4 |
| **Ice Cream Sandwich** (4.0, 4.03) | 16/12/2011 | 154 | 350 | 2.3 |
| **Jellybean** (4.1, 4.2, 4.3) | 24/07/2013 | 586 | 1,922 | 3.3 |
| **KitKat** (4.4) | 31/10/2013 | 99 | 781 | 7.9 |
| | | ∑ = 2,176 | ∑ = 5,354 | x̄ = 2.7 |

* Total number of requests logged between the first beta release on 16/11/2007 and Android version 1.1 released on 09/02/2009

classified, and to verify that misspellings and plural forms of nouns were treated appropriately by our program (flagging each as true or false). We computed reliability measurements from these actions using Holsti's coefficient of reliability [17] to evaluate our agreement. Overall, our reliability check revealed 90% agreement, and the remaining 10% of codes were resolved by consensus. This represents excellent agreement between coders. We provide our results next.

## 4. RESULTS

We present our results in this section. We present our results for RQ1 in the first subsection, before presenting those for RQ2 in the second subsection. We offer our pre-processed data for replication (or other) studies at https://goo.gl/4qLVcf.

*A. RQ1. Do Android developers respond to end-users' requests for application improvements logged via the issue tracker?*

We examine Google's responses to end-users' requests and in doing so extend our previous work [7]. In Licorish et al. [7] the top most problematic Android OS features were identified (i.e., those most often mentioned in enhancement requests), and analyses revealed how complaints about these features varied over time and how complaints were interconnected. In the current work we use the same top 20 feature enhancement requests identified in our previous work to check how Android development teams responded to those requests. Fig. 1 shows the top 20 features mentioned, and demonstrates that contacts, screen, notification, call and calendar were the top five features in enhancement requests across all versions. The average count for the top 20 is 147.6 requests per feature.

---

[5] http://www.merriam-webster.com/dictionary/noun

[6] http://www-01.sil.org/linguistics/GlossaryOfLinguisticTerms/WhatIsANoun.htm

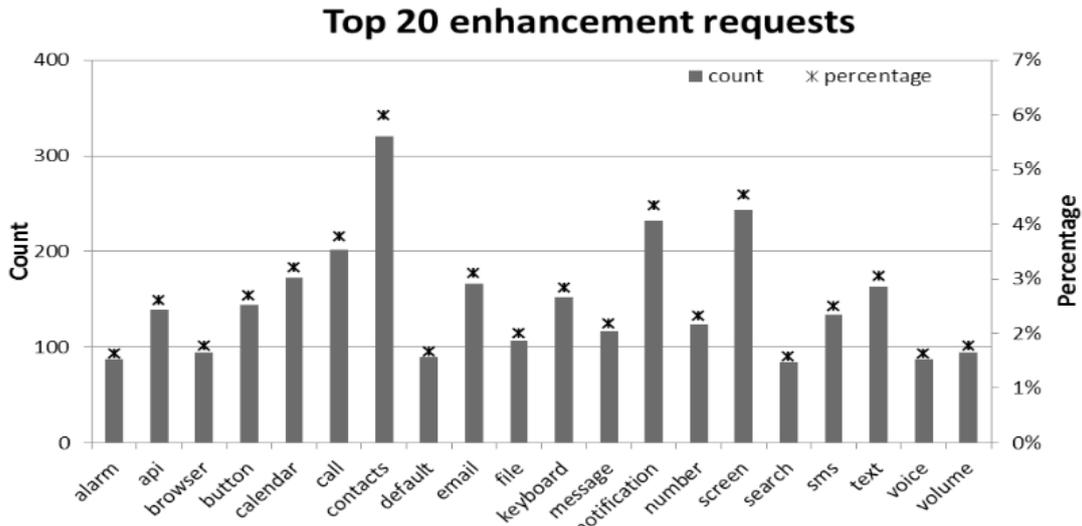

Fig. 1. Android's top 20 most frequent enhancement requests.

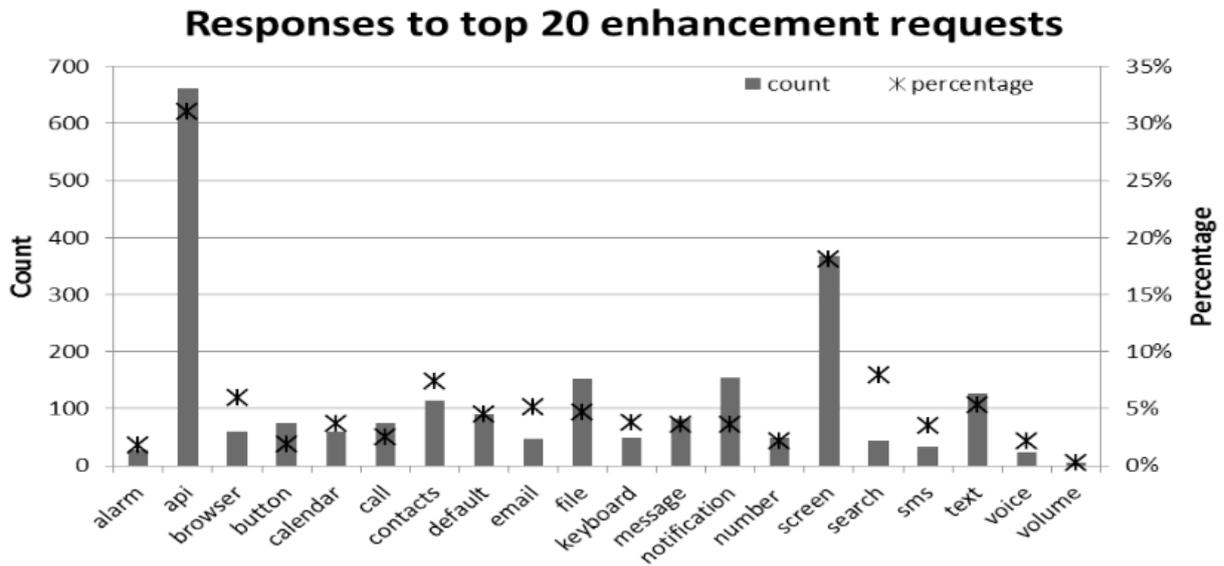

Fig. 2. Response to Android's top 20 most frequent enhancement requests.

We next examined Google's responses to these requests in Fig. 2, which shows the response count and percentage for each requested feature in the top 20. In total, there are 2,300 responses to the top 20 enhancement requests in all release notes, with an average of approximately 115 responses per feature (median= 67, Std Dev= 498.8). We observe that *api* and *screen* were the most frequently mentioned features (of those in the top 20 feature enhancement requests) in the Android OS official release notes across all versions. We further observe that *api* was mentioned 661 times in total, whereas *screen* was mentioned some 368 times. These are followed by *notification* (155), *file* (152), *text* (127) and *contacts* (115). On the other hand, *volume* appears to be the least frequently mentioned feature, arising only six times (0.27%) in the release notes. By taking the average of the top 20 responses as a threshold, we found that there were six features with a response value equal to or higher than the average (i.e., *api*, *screen*, *notification*, *file*, *text* and *contacts*).

Overall, in examining requests and responses it is notable in Fig. 1 and Fig. 2 that screen and notification are both prominent in the top 5 requests and responses. While contacts, call and calendar appear to be less prominent in Fig. 1 and Fig. 2 (and compared to the top responses listed above), these did feature consistently in requests and responses. We triangulated our observations and investigated the relationship between feature enhancement requests by users and responses by Android development teams to see if there was significant correlation between requests and responses. We first assessed the normality of data using the Shapiro-Wilk test, finding that the data distribution violated normality. We thus selected the non-parametric Spearman's rho ($r$) correlation test for our triangulation. The degree of association between variables is assessed using Cohen's classification [18]: it is interpreted that there is a low correlation when $0 < r \leq 0.29$, medium when $0.3 \leq r \leq 0.49$ and high when r ≥ 0.50. This interpretation also applies to negative correlations, although the association is inverse rather than direct [19].

TABLE II. SPEARMAN'S CORRELATION BETWEEN FEATURE REQUESTS AND RESPONSES OVER ANDROID VERSIONS

| Feature | r | p-value | Feature | r | p-value |
|---|---|---|---|---|---|
| Early | 0.23 | 0.32 | Gingerbread | 0.00 | 0.99 |
| Cupcake | 0.18 | 0.45 | Honeycomb | 0.40 | 0.08 |
| Donut | -0.08 | 0.73 | **ICS** | **0.49*** | **0.03*** |
| Éclair | 0.21 | 0.39 | **Jellybean** | **0.57*** | **0.01*** |
| Froyo | 0.22 | 0.34 | KitKat | 0.27 | 0.26 |

TABLE III. SPEARMAN'S CORRELATION BETWEEN FEATURE REQUESTS AND RESPONSES FOR INDIVIDUAL FEATURES

| Feature | r | p-value | Feature | r | p-value |
|---|---|---|---|---|---|
| alarm | 0.23 | 0.53 | keyboard | 0.34 | 0.34 |
| api | 0.30 | 0.39 | **message** | **0.66** | **0.04*** |
| browser | 0.14 | 0.69 | notification | 0.55 | 0.10 |
| button | 0.53 | 0.11 | **number** | **0.77** | **0.01*** |
| calendar | 0.17 | 0.64 | screen | 0.46 | 0.19 |
| **call** | **0.69** | **0.03*** | search | -0.55 | 0.10 |
| contacts | 0.07 | 0.85 | sms | 0.22 | 0.53 |
| default | 0.48 | 0.16 | **text** | **0.71** | **0.02*** |
| email | 0.09 | 0.81 | voice | 0.16 | 0.67 |
| file | 0.42 | 0.23 | **volume** | **0.79** | **0.01*** |

For all statistical tests we assume a significance threshold of 5% (i.e., p ≤ 0.05).

Our test considered the potential association between the total numbers of requests and responses for each feature in the top 20 requested features. We found that there was a statistically significant and strong correlation between the number of users' requests and the number of Google's responses to these requests (r=0.54, p-value= 0.01). This result suggests that the Android development team generally responded to the top enhancement requests made by the Android community. We next examine this pattern of requests and responses over the Android OS lifetime.

*B. RQ2. How are end-users' requests addressed overtime?*

Fig. 3 and Fig. 4 depict the numbers of enhancement requests and responses to those requests for all the above-average features across OS releases, respectively (i.e., how end-users' requests were addressed over time). It is evident in Fig. 4 that few responses to enhancement requests were enacted in the Early versions. Responses generally started to increase from Cupcake. Particular features received a high number of responses to users' requests in specific versions. For example, there is a marked increase in responses to *api* requests from Donut to Éclair, and then from Gingerbread to Honeycomb. In addition, there was a notable increase in the consideration of *screen* requests from Gingerbread to Honeycomb. It is also important to highlight that the responses to feature requests decreased markedly from Jellybean and KitKat for all those six features in Fig. 4. This mirrors decreases in the above average feature requests between Jellybean and KitKat as shown in Fig. 3. Thus, the general trend in the two graphs is one of convergence.

We compared those features that appeared in both graphs to further examine request-response convergence (i.e., *screen*, *notification*, *text* and *contacts*). As is shown in Fig. 3 and Fig. 4, screen was given the most attention between the release of Gingerbread and Honeycomb, whereas the highest number of requests for this feature was made during Ice Cream Sandwich (ICS) and Jellybean. Thus, Google developers appeared to pay more attention to concerns about the screen feature in the earlier versions, even though this issue was not the most pressing of those raised at that time (see trends for *contacts*, *call* and *notification* in Fig. 3). This pattern was the opposite of that noted for notification, text and contacts, which saw request-response convergence between ICS and Jellybean.

In fact, in looking at the general trends of requests in Fig. 3 we observe a uniform increase in users' requests between Froyo and Gingerbread for all features, which then reduced before increasing again between ICS and Jellybean. On the other hand, although responses increased somewhat between Froyo and Gingerbread, and ICS and Jellybean in Fig. 4, there were also other instances of high levels of Google responses, with the overall pattern being quite variable for the different features.

To triangulate our observations in Fig. 3 and Fig. 4 we formally examined the correlations between enhancement requests and responses across all Android OS major versions (as detailed in Table I). The results of our correlation analysis are reported in Table II: all statistically significant outcomes are marked with a star (*) and strong and medium correlations are shown in **Bold** typeface. Examining these correlation results it is evident that requests and responses are significantly correlated for two of the ten versions (i.e., for ICS and Jellybean). These correlations are strong for the Jellybean releases *(r=0.57 and p-value=0.01)*, and medium over ICS releases *(r=0.49, p-value=0.01)*.

We next investigated which particular features (of the top 20 feature enhancement requests) were most often addressed by the Android development team. We thus conducted correlation tests examining the relationship between requests made and responses provided for each of the top 20 features introduced in Fig. 1 and Fig. 2, for all the Android versions.

Table III reports the correlation analysis for our tests, again, showing how feature enhancement requests and responses to these requests for each individual feature were related. As shown in Table III, we found that five of the top 20 features were observed to have a statistically significant and strong association between requests and responses. Feature enhancement requests and responses were strongly correlated for the following features: *call (r=0.69, p-value=0.03)*, *message (r=0.66, p-value=0.04)*, *number (r=0.77, p-value=0.01)*, *text (r=0.71, p-value=0.02)* and *volume (r=0.79, p-value=0.01)*. The following section discusses these results (and those above) and considers the implications of our findings for the OADP community.

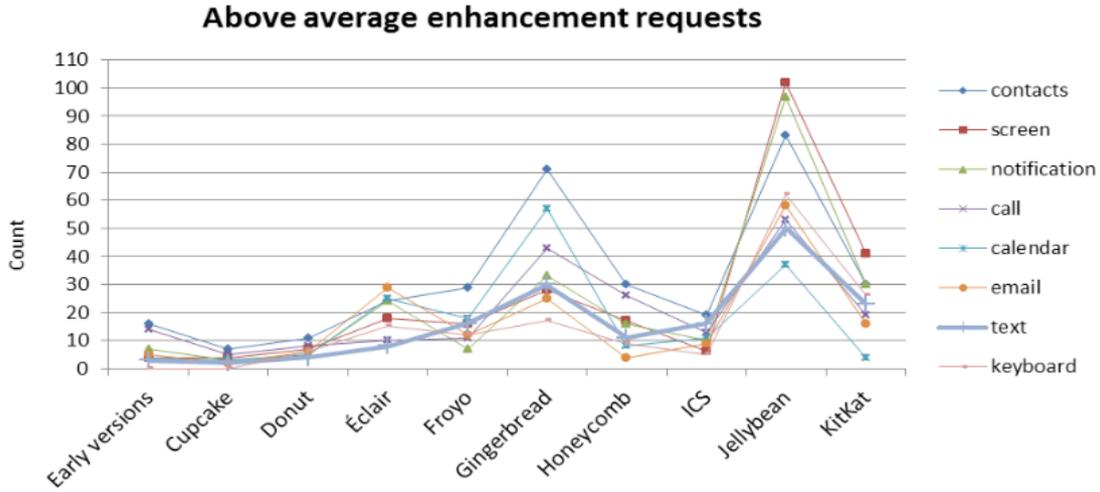

Fig. 3. Above average features for top 20 enhancement requests.

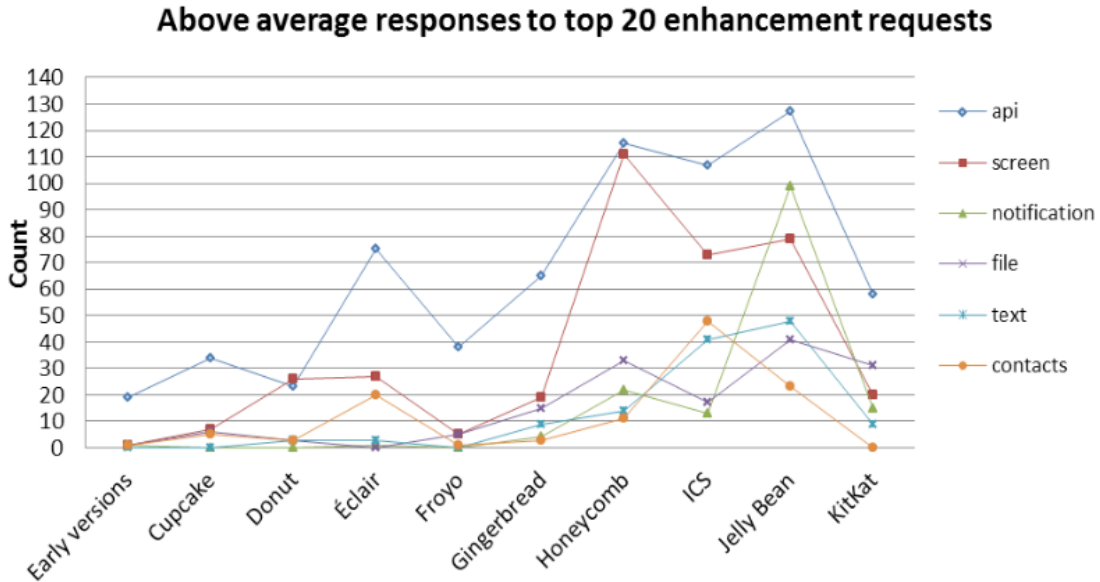

Fig. 4. Above average responses for top 20 enhancement requests.

## 5. DISCUSSION AND IMPLICATIONS

We discuss our findings in this section. We examine our outcomes to answer RQ1 in the first subsection. We then analyze our findings in answering RQ2 in the second subsection. Finally, we consider the implications of our outcomes in the third subsection.

*A. RQ1. Do Android developers respond to end-users' requests for application improvements logged via the issue tracker?*

**The Android development team generally responded to the top enhancement requests made by the Android community.** Our results show that there was a strong correlation between end users' requests for enhancement and Android developers' responses to those requests. This evidence suggests that, even in environments where a large number of requests are forwarded to developers, those *developers may still achieve the balance* that is typically seen in conventional settings were end-users are involved during the system development life cycle, and particularly in early stages [7]. We thus believe that *user-centered/focused design is still necessary for OADPs* and in environments where MDRE is used as a strategy for evolving a product. In fact, we contend that our findings also provide lessons for individuals operating in those settings regarding the feasibility of prioritizing requirements in order to satisfy previously unknown user groups [2]. We see a strategy that focuses on addressing users' top concerns as very useful, both for satisfying users and potentially maintaining competitiveness. In fact, notwithstanding the challenges associated with requirements overload, requirements ambiguity and prioritization, high numbers of end-users' reviews and requests may appropriately inform product quality [2]. Thus, there is utility in efforts that aim to extract the most important users' concerns [10], and the most relevant and noteworthy requirements [8] evident in large numbers of end-users' reviews/complaints.

Our results show that, for the Android OS, ***contacts, screen, notification, call*** and ***calendar*** were the top five features requested for enhancement, whereas ***api***, ***screen***, ***notification***, ***file***, and ***text*** were most prominent in Google's responses. We also observed that, in particular, ***api*** was mentioned frequently in the Android release notes (Fig. 2).

We suspect that this may be due to strategic development of APIs by Android teams for various applications and platforms. APIs are the main interaction channel between software applications. Increases in the number and breadth of features and applications are likely to be mirrored by an increase in the development of APIs for those features and applications. This feature (*api*) also groups various Android requests (e.g., file storage api, camera api, and so on), and thus, it could be expected to be ranked higher than those that reflect a single functionality.

We triangulated our findings here by examining the features that were given most attention by Android teams, irrespective of end-users' requests. We thus examined the features mentioned in the Android official release notes. Similar to the analysis conducted on the feature requests (refer to Fig. 1), we extracted and aggregated the features mentioned in all release notes, and then selected the top 20 most frequently mentioned features. Fig. 5 shows the count and percentage of each feature in the top 20 features list. As shown in Fig. 5, *api* was the most frequently mentioned feature followed by *screen* and *data*. Other features such as "widget", "notification", "image", "camera", "text", "video" and "contact" also figured prominently. In fact, comparing the features that appear in the top 20 list in Fig. 5 (for responses) with those that were identified in the top 20 feature requests list in Fig. 1, we observed that there was a 45% overlap: nine of the top 20 frequently delivered features (as per the content of release notes) were mentioned in the top 20 feature requests by end-users. *This evidence suggests that end-users' feedback played an integral role in Android OS evolution, and these developers' portfolios.*

### B. RQ2. *How are end-users' requests addressed overtime?*

**We observed that while Android developers provided few responses to users' requests over the early versions of the Android OS, this trend changed as the software evolved.** We anticipated that requirements for early versions of the Android OS were likely driven by internal Google teams and other market research, thus, this result was expected. We further noted that some particular Android OS features received higher responses to users' requests in specific versions than others. In fact, *screen* and *api* issues were given the greatest amount of attention over the Android OS releases, with *api* responses being particularly pronounced (refer to Fig. 4). This evidence suggests that the latter releases of Android have indeed benefitted from user reviews, and that *specific issues were considered in developers' prioritization strategies*. While *screen* issues could indeed be expected to be given high priority, as it is an end-user facing feature and issues would likely affect a large cohort of users, it is insightful that issues in the lower *api* layers were also considered central to Android OS developers' feature enhancement strategy. This latter type of feature would typically reside in the application frameworks and libraries. Thus, evidence here suggests that those developing apps for the Android OS also employed the issue tracker to express their requests for changes. *This highlights the particular opportunity afforded by OADPs where a range of user types are able to express their concerns or desires in an effort to improve and extend previously deployed software [7].*

We observed that responses to feature requests decreased markedly after the release of Jellybean and KitKat (refer to Fig. 4), the latest versions in Table I. This trend was also seen for enhancement requests between Jellybean and KitKat in Fig. 3. (We should note, however, that our dataset only comprised a subset of the issues that were logged since the release of KitKat (from 31/10/2013 to 20/03/2014), which could have affected the pattern of results noted.) Overall, some issues were given more attention earlier than others, although these were not always central to users' requests. For instance, Google developers paid more attention to *screen* complaints in the earlier versions, although *contacts*, *call* and *notification* issues were most pressing. While these issues may be perceived as equally important, perhaps rework aspects of the screen were related to the release of specific (evolving) hardware, and hence, were given priority over other issues. There has been rapid growth in the evolution of Android- supported mobile hardware [20], which would no doubt affect developers' development and deployment strategies.

We observed a uniform increase in enhancement requests over the lifetime of the Android OS (refer to Fig. 3), suggesting that some releases were more problematic than others. In fact, our earlier work also detected this pattern for issues logged overall and for security issues [21]. In contrast here, however, and as noted above, developers did not always respond to the most pressing issues. Of the ten major Android OS releases examined, our results show that developers paid most attention to requests made between Honeycomb and ICS and ICS and Jellybean (refer to Table II). We observed that while there were few requests between Honeycomb and ICS, the opposite was seen between ICS and Jellybean. Thus, the attention given to improvements on issues over the latter range of releases was fitting; however, it is not entirely clear why the large amount of attention was directed towards issues raised over the former range of releases.

In terms of the attention given to specific features, overall, enhancement requests for ***call, message, number, text*** and ***volume*** were given particular attention over Android OS releases (refer to Table III) – such issues were singled out for attention over others. Importantly, these features are end-user facing, and are perhaps used the most by end users. This evidence has implications for OADP communities, in terms of prioritizing the most important user requests from a voluminous pool of reviews. Beyond top complaints, prioritization strategies for developers of apps on OADPs and by those that employ MDRE approaches may also consider frequency of feature use. We further consider this issue in the next subsection.

### C. Implications for Text Mining and Software Engineering

*From a theoretical perspective, techniques aimed at extracting requirements from end users' reviews in OADPs and portals that capture MDRE offer a useful avenue for developers to provide user-centered software deliverables.* Text mining approaches are considered central to such an endeavor. Research has indeed examined the feasibility of these approaches, providing promising results about their potential utility [22-23]. Here we provide further demonstration of the utility of such an approach in

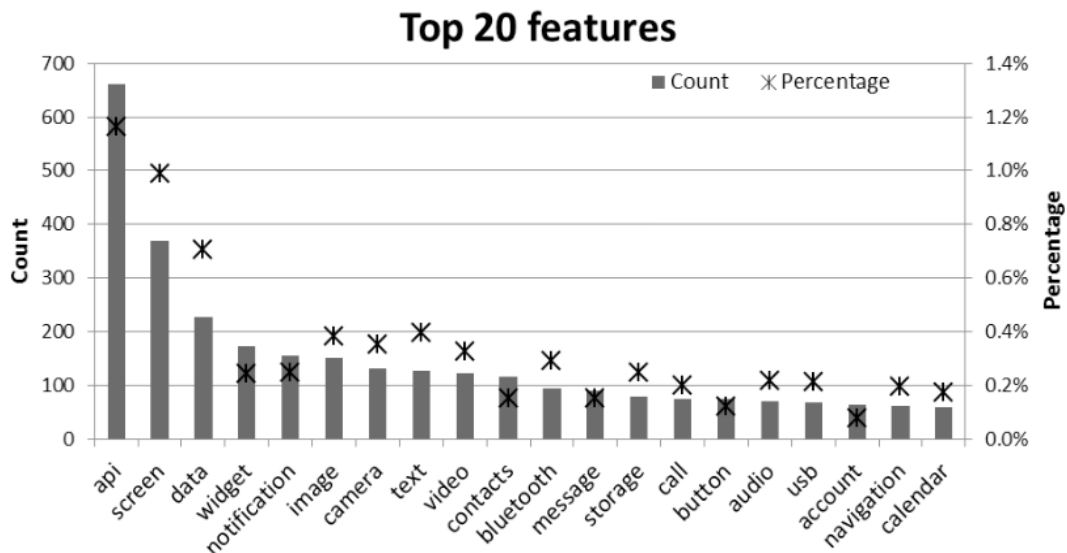

Fig. 5. Android's top 20 most frequent features.

extracting knowledge for the software engineering community. Based on our results we believe that continued efforts spent advancing research work aimed at improving the accuracy and precision of these methods would be useful for both the text mining and software engineering communities.

In fact, our outcomes in this work have several implications for software development practitioners, and for companies whose software are provided on OADPs. *We observe that users contribute meaningfully to the feedback loop in improving software in OADPs through their reviews and the avenues they identify for improving software [7]*. Thus, these stakeholders should be regarded as such. Although voluminous, developers should consult users' reviews to identify ways to improve their offerings, and to remain competitive [2]. *Our evidence indicates that, even in these new development contexts, the established wisdom that involving end-users during system development, and particularly in early stages, provides critical insights into the likely context of system use and will aid with product performance, remains supported*. Such a user-centered approach is considered necessary for traditional modes of software development (e.g., agile approaches). We would assert that, although there are challenges related to prioritizing requirements in OADP environments, and especially in order to satisfy previously unknown user groups with the first release of a product, continual releases should embrace users as being part of the software evolution process. Thus, efforts should be invested to capture and consider their views. Such a move is likely to have a positive impact on software product outcomes.

While it may well be infeasible to address all of the concerns raised by end users online in OADPs, given the potentially voluminous number of reviews, *a strategy aimed at addressing the most pertinent concerns would go some way to assuring users that their views are being sought and responded to*. For instance, in the consideration of a GPS app, issues related to signaling and battery life may be critical to a user-base, and thus, if assessed as such should attract developers' responses. Similarly, developers of a game app may find that efforts aimed at optimizing battery (or other) resource use could contribute to maintaining users' interest. Developers should thus strategize between both pressing users' concerns and the nature of the service that is provided to users. Evidence of the former is easily detected in end users' reviews.

## 6. THREATS TO VALIDTY

While we have examined an important topic area, and have provided insights into the usefulness of end-users in the feedback loop for improving product performance in OADPs, there are shortcomings to this work that may affect its generalizability. We consider these in turn.

We acknowledge that the Android OS community might not be thought of as a typical OADP, where developers commonly consist of individuals or small teams that have limited resources and may not be able to conduct extensive tests of their product before launch. Instead, the Android OS is sponsored primarily by Google, whose substantial resources may be employed for UAT, and thus, this product may be of very good quality. However, given that the Android issue tracker brings together developers and users, such that users can provide diverse feedback on this product's performance and improvement opportunities, we believe that lessons learned from such a case would be relevant to OADPs.

Although the Android issue tracker is publicly hosted, and so is likely to capture most of the community's concerns [24], issues may also be informally communicated to and addressed within the development teams at Google.

Similarly, unreported issues are not captured by our analysis. We also accept that there is a possibility that we could have missed misspelt features. That said, the convergence of our results which examined multiple separate data sources (issue logs and release notes) suggests that our approach was generally robust. In fact, our reliability assessment measure revealed excellent agreement between coders, suggesting that our findings benefitted from accuracy, precision and objectivity [17].

We separated issues based on the dates of the major Android OS releases. Given that device manufacturers have been shown to delay upgrading their hardware with recent Android OS releases [25], there is a possibility that some issues reported between specific releases were in fact related to earlier releases. However, this misalignment was not detected during our previous contextual analysis [21], suggesting that our approach appropriately classified issues.

Finally, although the issue trackers of many mobile OSs are not publicly available, and the distribution of these OSs' issues may not be similar to what is observed in this work for the Android OS, mobile OSs such as Microsoft Windows, Apple iOS, Symbian and BlackBerry are all likely to follow release-maintenance cycles similar to that of Android OS in order to remain competitive in the market.

## 7. CONCLUSIONS

End-users' involvement is integral to software product performance outcomes, and particularly when users' views are central to the features that developers provide. However, as highlighted throughout this paper, users' views are not conventionally accessible in OADPs and in MDRE contexts, where software is developed for many potential users and where the traditional requirements gathering process, and the negotiation of requirements with a single group of customers do not apply. Rather, in these contexts software is released to previously unknown user groups, who then provide reviews on the product's features and their performance. Thus, there are often challenges in terms of how these (potentially large numbers of) reviews are managed in the decision-making process of gathering and prioritizing improvement suggestions or requirement requests. Given that reviews provide rich content comprising users' opinions and preferences for a software product, this source can also support the evolution of a software product. We have indeed demonstrated this utility in this work. Moreover, we have attempted to address the gap of reconciling developers' responses with users' reviews, using Android OS as our case study. In addition, we have also validated the utility of text mining and NLP techniques for aiding this cause.

Our evidence shows that the Android development team generally responded to the top enhancement requests made by the Android community, suggesting that user-centered/focused design is still necessary for OADPs and in environments where MDRE is used as a strategy for evolving a product. We also observed that while Android developers provided few responses to users' requests over the early versions of the Android OS, this trend changed as the software evolved, with particular features receiving more attention than others. Furthermore, our evidence shows that enhancement requests logged (and Google's responses) related to various layers of the Android OS, suggesting that a range of users logged issues online. Opportunities to engage with such a diverse range of users are not readily afforded in more conventional forms of software development where a single group of users tend to inform requirements and/or perform software testing.

We believe that efforts aimed at increasing the accuracy and precision of text mining techniques would be useful for the text mining and software engineering communities. In addition, developers are encouraged to view OADPs and MDRE platforms as valuable sources of product feedback. In particular, listening to the most pressing end-user feedback delivered via such platforms may improve their satisfaction and aid product performance. We plan to examine this latter issue in our future work, in particular, we are interested in understanding why some end-users' requests were given more attention than others.

## ACKNOWLEDGMENT

Thanks to Google for making the Android issues and release notes publically accessible to facilitate the analyses that are performed in this study.